\documentclass[9pt,letterpaper,twocolumn]{extarticle}

\pdfoutput=1

\usepackage{soul,enumerate}
\usepackage{comment}
\usepackage{titlesec} 

\usepackage{dsfont}
\usepackage{graphicx}
\usepackage{subcaption}
\usepackage[margin=0.9in]{geometry}
\usepackage[usenames,dvipsnames]{color}
\usepackage[colorlinks,linkcolor=Blue,urlcolor=Blue,citecolor=Blue]{hyperref}
\usepackage{amsmath,amssymb}
\usepackage{amsfonts}
\usepackage{bbm}
\usepackage[squaren]{SIunits}

\usepackage{etoolbox}
\apptocmd{\thebibliography}{\setlength{\itemsep}{-4pt}}{}{}

\usepackage{mathpazo} 
\usepackage{courier} 
\normalfont

\usepackage[T1]{fontenc}
\usepackage{caption}
\usepackage{upgreek}

\newcommand{\ad}[1]{\textsuperscript{#1}\kern-2pt}

\usepackage{amsmath,amsthm,mathtools}

\def\({\left(}
\def\){\right)}
\def\[{\left[}
\def\]{\right]}

\makeatletter
\def\blx@maxline{77}
\makeatother

\usepackage{capt-of}

\usepackage[ruled]{algorithm2e} 
\usepackage{newfloat,algcompatible} 
\def\({\left(}
\def\){\right)}
\def\[{\left[}
\def\]{\right]}  
  
\def\mytitle{ 
Ultra-compact integrated graphene plasmonic photodetector with bandwidth above 110\,GHz
\vspace{-4mm}}  

\title{\vspace{-1.0cm}\huge\textbf{\textrm{\mytitle}}}  
 
\author{Yunhong Ding$^{1, 2}$*, Zhao Cheng$^{1, 6}$, Xiaolong Zhu$^{5}$, Kresten Yvind$^{1, 2}$, Jianji Dong$^{6}$, Michael Galili$^{1, 2}$, Hao Hu$^{1, 2}$,\\
N. Asger Mortensen$^{3,4}$, Sanshui Xiao$^{1, 4}$, Leif Katsuo Oxenl{\o}we$^{1, 2}$}

\date{} 
\begin{document}
\twocolumn[{
\maketitle 
\vspace{-8mm}
\begin{center}
\begin{minipage}{0.95\textwidth}
\begin{center}
\textit{\textrm{
\textsuperscript{1} 
Department of Photonics Engineering, Technical University of Denmark, 2800 Kgs. Lyngby, Denmark
\\\textsuperscript{2} 
Center for Silicon Photonics for Optical Communication (SPOC), Technical University of Denmark, 2800 Kgs. Lyngby, Denmark
\\\textsuperscript{3} 
Center for Nano Optics and Danish Institute for Advanced Study, University of Southern Denmark, Campusvej 55, DK-5230 Odense M, Denmark
\\\textsuperscript{4} 
Center for Nanostructured Graphene (CNG), Technical University of Denmark, DK-2800 Kongens Lyngby, Denmark
\\\textsuperscript{5} 
Department of Micro- and Nanotechnology, Technical University of Denmark, DK-2800 Kongens Lyngby, Denmark
\\\textsuperscript{6} 
Wuhan National Lab. for Optoelectronics, Huazhong University of Science and Technology, 430074, Wuhan, China.\\
*Emails: 
\textcolor{blue}{yudin@fotonik.dtu.dk}
\\
}}
\end{center}
\end{minipage}
\end{center}

\setlength\parindent{12pt}
\begin{quotation}
\noindent 
{
\textbf{Graphene-based photodetectors, taking the advantages of high carrier mobility
and broadband absorption
in graphene, have recently experienced rapid development.
However, their performance with respect to responsivity and bandwidth is still limited by weak light-graphene interaction and large resistance-capacitance product. Here, we demonstrate a waveguide coupled integrated graphene plasmonic photodetector on a silicon-on-insulator platform. Benefiting from plasmonic enhanced graphene-light interaction and subwavelength confinement of the optical energy, we achieve a small-footprint graphene-plasmonic photodetector working at the telecommunication window, with large bandwidth beyond 110\,GHz and high intrinsic responsivity of 360\,mA/W. Attributed to the unique electronic bandstructure of graphene and its ultra-broadband absorption, operational wavelength range extending beyond mid-infrared, and possibly further, can be anticipated. Our results show that the combination of graphene with plasmonic devices has great potential to realize  ultra-compact and high-speed optoelectronic devices for graphene-based optical interconnects.}
}
\end{quotation}}]


The fast development of silicon photonics makes it feasible to construct optical interconnects that can replace electrical interconnects for chip-level data communications with low-energy consumption and large bandwidth~\cite{KCS1982,DAM2009}. However, for photodetection in the silicon-based optical interconnect, it still needs to be integrated with another absorbing material, e.g., germanium or III-V compound semiconductor~\cite{MJ2010,IH2004}, leaving a big challenge for direct monolithic integration with the complementary metal-oxide-semiconductor (CMOS) technology and in achieving high bandwidth limited by absorbing materials' poor electrical properties. Graphene, a unique CMOS compatible two-dimensional (2D) material, provides great potential in the realization of high-performance optoelectronic devices~\cite{ML2011,VS2018,YD2017,TG2012,SY2017,ZS2010}. In particular, significant efforts have been devoted to graphene photodetectors (PDs)~\cite{TM2010,XG2013,AP2013,XW2013,DS2014,CL2014,SS2016,IG2016,MP2018arxiv}. The distinct properties of graphene in terms of ultrahigh carrier mobility~\cite{KB2008,VD2010}, zero bandgap property that enables wavelength-independent light absorption over a very wide spectral range~\cite{RN2008,JD2008,SCakmakyapan2018}, and tunable optoelectronic properties~\cite{ZL2008,FW2008} give rise to realize graphene photodetectors with large spectral bandwidth and high speed.  

Graphene PDs rely on devices with broken inversion symmetry~\cite{TM2010,XG2013,FK2014}.
For graphene-based integrated PDs, the inversion symmetry can be conveniently relaxed through an asymmetric positioning of the waveguide with respect to the graphene coverage, and several devices ~\cite{XG2013,RJS2015} have been reported. However, due to the modest light-matter interaction between the single-layer graphene (SLG) and the waveguide mode, the size of devices has to be at least tens to hundreds of microns to achieve a reasonable responsivity, thus limiting high-speed operation. A small device however gives weak absorption of light, and thus eventually low responsivity. This counter-acting effect represents a big challenge for graphene-based PDs supporting both high-responsivity and large bandwidth. So far, the state-of-the-art bandwidth of the graphene PD is $\sim$76\,GHz, however with a modest responsivity of 1\,mA/W~\cite{DS2017}. Another promising scheme is to break the symmetry of potential profile of the device by different metallic-induced doping~\cite{GG2008} near metal-graphene contact regions. In this scheme, an internal (built-in) electric field~\cite{KN2004} is formed to separate photo-generated carriers~\cite{FX2009,TM2010}. A milestone for high-speed photodetectors based on SLG has been demonstrated by free-space top-illumination technique~\cite{TM2010}. However, the internal built-in electric field only exists in narrow regions of $\sim$200\,nm adjacent to the electrode/graphene interfaces~\cite{FX2009,TM2010}. The large distance of 1\,$\mu$m between the two electrodes~\cite{TM2010} limits collection efficiency of photo-generated carriers and thus the responsivity. Moreover, the required very high-quality graphene typically relies on the exfoliation method~\cite{KN2004}, which restricts the potential for large scale integration.

Here, we report an ultra-compact, on-chip, and high-speed graphene photodetector based on a plasmonic slot waveguide~\cite{MA2017,YD2017,YS2018}. The subwavelength confinement of the plasmonic mode gives rise to the enhanced light-graphene interactions, and the narrow plasmonic slot of 120\,nm enables short drift paths for photogenerated carriers. A smallest integrated graphene photodetector with a graphene-coverage length of 2.7\,$\mu$m is demonstrated without response dropping up to a frequency of 110\,GHz and an intrinsic responsivity of 25\,mA/W. Increasing the device size to 19\,$\mu$m results in an increased intrinsic responsivity of 360\,mA/W, equivalent to a high external quantum efficiency of 29$\%$. These performances are comparable with a state-of-the-art commercial 100\,GHz semiconductor photodetector~\cite{XPDV412xR}, and can further be improved significantly. With the extremely broad absorption band of graphene, our device shows great potential. Moreover, the use of chemical-vapour-deposition(CVD)-growth graphene here allows for the scalable fabrication and we believe that our work greatly pushes the 2D material towards practical applications, e.g. in optical interconnects, high-speed optical communications, and so on.

\begin{figure}[t!] 
\centering
\includegraphics[width=0.46\textwidth]{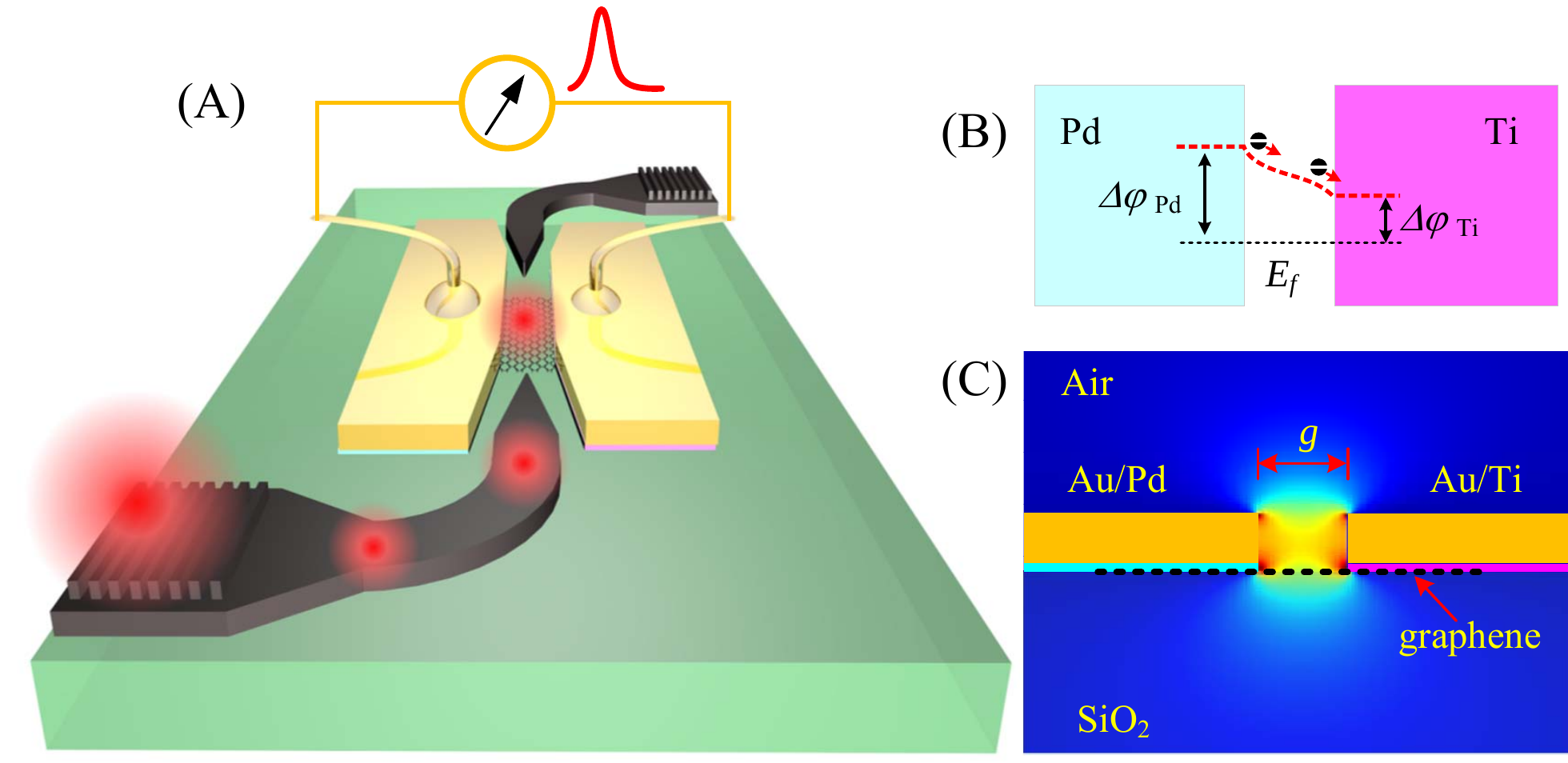} 
\caption{\textbf{Principle of the graphene-plasmonic integrated photodetector.}  
\textbf{A}. Schematic of the proposed graphene-plasmonic hybrid photodetector. \textbf{B}. The potential profile of the device showing the drift of the photo-generated carrier. $\Delta\phi_{\rm Pd}$ and $\Delta\phi_{\rm Ti}$ are the difference between the Dirac point energy and the Fermi level in palladium- and titanium-doped graphene, respectively. 
\textbf{C}. Cross-section of the device with its corresponding plasmonic slot mode at $\lambda=1.55~\mu m$. 
}
\label{fig:principle}
\end{figure}  

\section*{Results}

\subsection*{Principle}

The schematic of the proposed graphene plasmonic hybrid photodetector is shown in 
Fig.~\ref{fig:principle}(A), where the light from a fiber is first coupled to a silicon waveguide through a grating coupler, and further to the plasmonic slot waveguide by a short taper structure~\cite{YD2017,YS2018}. 
The plasmonic slot waveguide, see the cross-section of the device in Fig.~\ref{fig:principle}(C), consists of two asymmetric metallic contacts of Au(90\,nm)/Pd(5\,nm) and Au(90\,nm)/Ti(5\,nm), resulting in different doping in the graphene~\cite{TM2010}. The potential difference indicated in  Fig.~\ref{fig:principle}(B) gives rise to the efficient separation of the photo-generated carriers and formation of photocurrent. The performance with respect to responsivity and speed of the proposed photodetector really takes the advantages of the plasmonic slot waveguide with the narrow gap $g$ of 120\,nm. Firstly, the plasmonic slot waveguide provides sub-wavelength light confinement in the nanometer-scale, as shown in Fig.~\ref{fig:principle}(C), resulting in extremely strong graphene-light interaction and thus high responsivity. With the optimum geometry of small gap and thin Au thickness of the plasmonic slot waveguide, the single-layer graphene leads to extremely high light absorption of $\sim$1\,dB/$\mu$m, see the Supplementary Material, which is at least more than one order of magnitude higher than that of graphene-silicon waveguide photodetectors~\cite{XG2013,DS2017}. Secondly, the internal electric-field mentioned above covers the whole plasmonic slot region of 120\,nm. Thus, photo-generated carriers can be effectively separated, leading to high responsivity. Furthermore, the narrow plasmonic slot gives short drift paths for carriers, resulting in an ultra-fast carriers transition through the photodetection region and thus high speed.  

\subsection*{Experimental results}

\begin{figure}[!t]
\centering
\captionsetup{width=0.46\textwidth}
\includegraphics[width=0.48\textwidth]{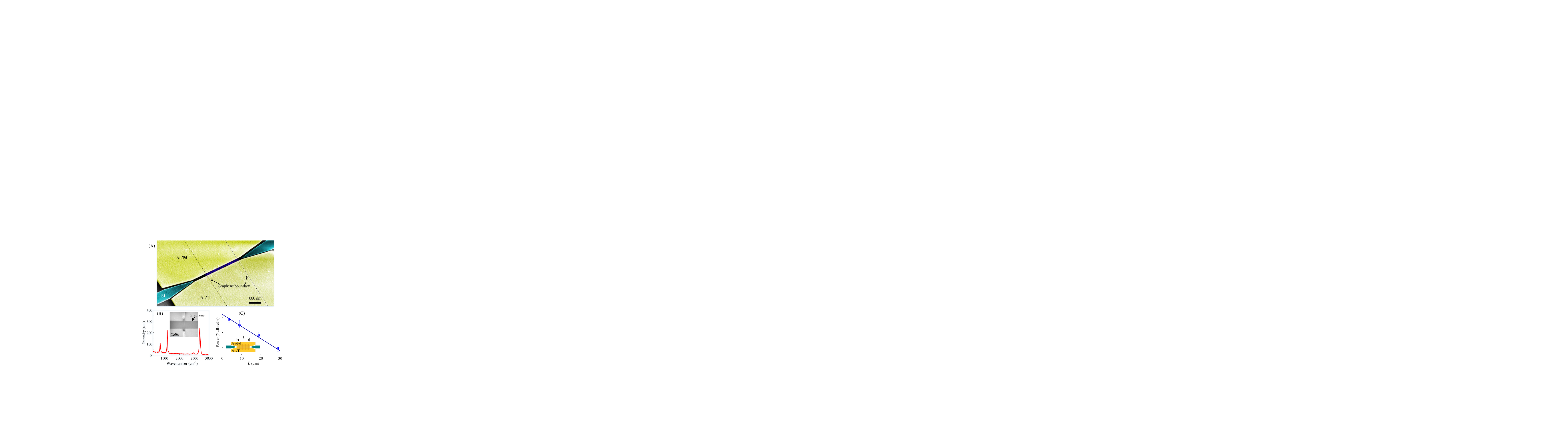}
\caption{\textbf{Characterization of the fabricated device.}
\textbf{A}. An example of fabricated graphene plasmonic PD with the graphene coverage length of 2.7\,$\mu$m. \textbf{B}. Measured Raman spectrum for the graphene after wet-transferring process. The inset shows the scanning-electron microscope (SEM) image of the device with 2.7\,$\mu$m after graphene patterning process. \textbf{C}. Analysis of coupling and propagation loss by the cut-back method. The error bar is obtained by measuring the six-copy device with the same graphene coverage length. } 
\label{fig:fabricated}
\end{figure}

Fig.~\ref{fig:fabricated}(A) shows the fabricated graphene-plasmonic photodetector, where the dashed lines represent the graphene coverage boundary. The Raman spectrum shown in Fig.~\ref{fig:fabricated}(B) illustrates moderate degradation after the wet-transferring process, which can be found in Method. Graphene plasmonic hybrid photodetectors with different graphene-coverage lengths were fabricated, and the cut-back method shows 
absorption coefficient of 0.8\,dB/$\mu$m in the detection region, as presented in Fig.~\ref{fig:fabricated}(C). 


\begin{figure*}[t!]
\centering
\captionsetup{width=0.85\textwidth}
\includegraphics[width=0.65\textwidth]{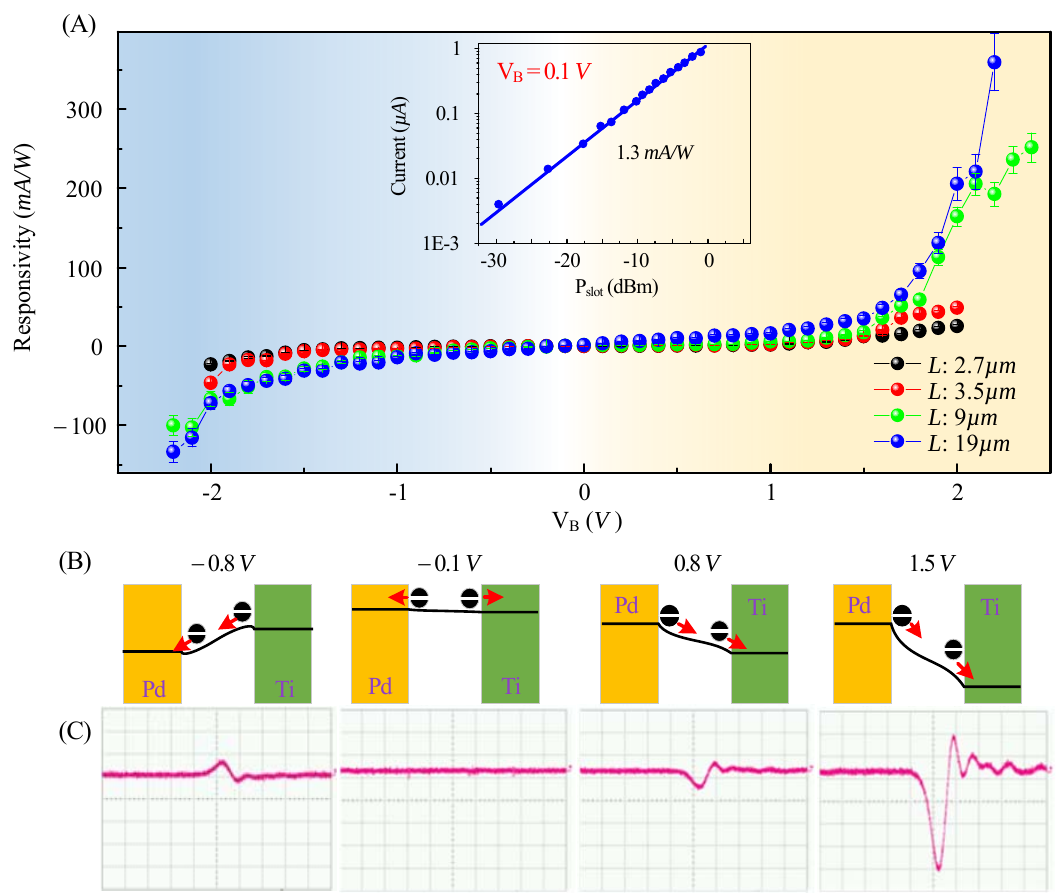}
\caption{\textbf{Characterization of responsivity of the devices.}
\textbf{A}. Measured responsivity at different bias voltage for four devices with different $L$. The inset illustrate the measured photocurrent as a function of the light power with zero bias.
\textbf{B}. Schematics of the potential profiles at the different bias. \textbf{C} Their corresponding measured output electrical signals for the photodetector with graphene-coverage length $L$ of 19\,$\mu$m. The time scale is 20\,ps/div.
} 
\label{fig:responsivity}
\end{figure*}


Being a key parameter, the intrinsic responsivity $R_{\rm ph}$ of the graphene-plasmonic waveguide photodetector with respect to the power in the plasmonic waveguide was characterized with light power of --4\,dBm (400\,$\mu W$) in the plasmonic waveguide. 
The photocurrent was measured 
by switching on and off the light, while recording the current difference. Measurements were performed at different bias voltage $V_B$ (Au/Ti electrode relative to Au/Pd electrode) for devices with different graphene-coverage length $L(=2.7, 3.5, 9, 19~\mu m)$, as shown in Fig.~\ref{fig:responsivity}(A). The potential profile at different bias voltage and their corresponding output electrical signals, due to injection of an optical pulse into the chip, are also presented in Fig.~\ref{fig:responsivity}(B) and ~\ref{fig:responsivity}(C), respectively. 
At the zero bias voltage, the intrinsic responsivity of $\sim$1.3\,mA/W
is observed for the device with the graphene-coverage length of 19\,$\mu$m, see the insets of Fig.~\ref{fig:responsivity}(A). It is due to the asymmetric potential profile between the two electrodes, which is also addressed in \cite{TM2010}. 
The zero responsivity measured at the negative bias around $-$0.1\,V indicates a flat potential profile as presented in Fig.~\ref{fig:responsivity}(B), and no change for the optical signal is observed, see the second subplot in Fig.~\ref{fig:responsivity}(C). Further increasing the negative bias voltage results in an increased responsivity, and a clear electrical pulse is obtained at $-$0.8\,V, as presented in Fig.~\ref{fig:responsivity}(C). Increasing the positive bias voltage also causes an increase of responsivity and a clear output electrical pulse with opposite polarity compared to the negative biasing case. 
A bias voltage beyond 1.5\,V leads to 
a significant increase of responsivity and much larger amplitude for the electrical pulse. Moreover, as the graphene-coverage length $L$ increases, the improvement for the responsivity is also observed, since a larger fraction of optical power is absorbed, thus in turn leading to higher photocurrent. As presented in Fig.~\ref{fig:responsivity}(A), the smallest device with 2.7\,$\mu$m graphene coverage gives the responsivity of $\sim$26\,mA/W at the bias voltage of 2\,V. The quantum yield represented by the external quantum efficiency (EQE) is defined by ${\rm EQE}=R_{\rm ph}\times \hbar\omega/e$, where $\hbar\omega$ is the photon energy, with $\hbar$ being the Planck constant and $\omega$ the frequency of light, while $e$ is the electron charge. Thus, the corresponding EQE of $2\%$ is obtained. For the device with 19\,$\mu$m graphene coverage, the highest responsivity of 360\,mA/W is obtained with the bias voltage of 2.2\,V, corresponding to a high EQE of $29\%$. 

\begin{figure*}[ht]
\centering
\captionsetup{width=0.8\textwidth}
\includegraphics[width=0.7\textwidth]{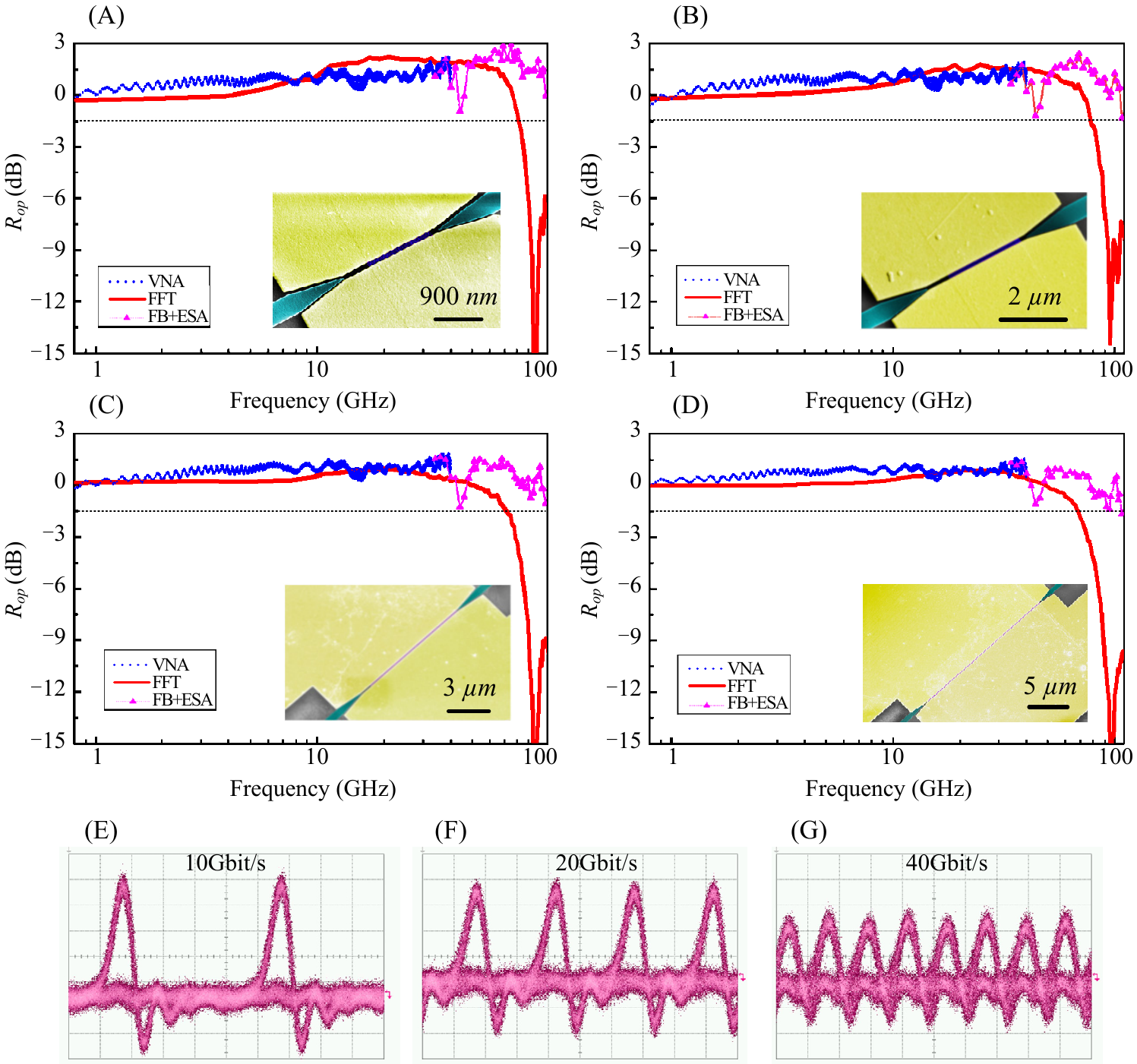}
\caption{\textbf{Bandwidth characterization of the devices.}
\textbf{A}--\textbf{D}. Measured optical bandwidth by VNA, impulse response, and frequency beating with ESA for the devices with graphene-coverage length of 2.7~$\mu$m, 3.5~$\mu$m, 9~$\mu$m, and 19~$\mu$m, respectively, at the bias voltage of 1.6\,V. The insets show the SEMs of associated fabricated devices.
\textbf{E}--\textbf{G}. Measured eye-diagram of RZ optical signal at 10, 20, and 40\,Gbit/s, respectively, for the device with 19\,$\mu$m graphene-coverage length biased at 1.6\,V.
} 
\label{fig:bandwidth}
\end{figure*}

We further measured the bandwidth of these graphene-plasmonic photodetectors, as presented in Figs.~\ref{fig:bandwidth}(A)--(D). The experimental setup can be found in the Supplementary Material. All measurements are carried out with the bias voltage of 1.6\,V to get a good signal-to-noise ratio. At the low frequency below 40\,GHz, we used a vector network analyzer (VNA, 40\,GHz bandwidth) to measure the  current (optical) frequency response $R_{\rm op}$. 
One can find no response-drop within the 40\,GHz bandwidth, see the blue lines in Figs.~\ref{fig:bandwidth}(A)--(D), indicating a large detection bandwidth. Then, the devices were further characterized by the alternative method of fast Fourier transformation (FFT) of the impulse response of the detectors. 
Within the 40\,GHz bandwidth, the frequency response obtained by the impulse-response method, see the red solid red lines, agrees quite well with the VNA method mentioned above. 
With the aid of the impulse response method, we obtain the measurement system optical bandwidth of at least 85\,GHz for the smallest device with graphene coverage length of 2.7\,$\mu$m. Increasing the device size results in broader impulse response (see the Supplementary Material) and thus smaller bandwidth. The device with graphene coverage length of 19\,$\mu$m exhibits the measurement system optical bandwidth of at least 75\,GHz. All the impulse response measurements for devices can be found in the Supplementary Material. Note that such systematical bandwidth is significantly influenced by the bandwidth of the oscilloscope, RF cable, and Bias Tee. Thus, we have employed the third method (FB+ESA) utilizing frequency beating (FB) between two coherent light fields to measure the RF power in wide-band electrical spectral analyzer (ESA, 110\,GHz bandwidth). 
As presented in Figs.~\ref{fig:bandwidth} (A)-(D), for the frequency near 40\,GHz, the response obtained by the FB+ESA method (the triangular marks) overlaps quite well with those by the VNA method (the blue curves). Within 110\,GHz, no drop in detection response is observed for the device with graphene coverage length of 2.7\,$\mu$m, indicating a bandwidth >110\,GHz. A longer device leads to slight drop in detection response at high frequency around 100\,GHz. The small frequency dip at $\sim$46\,GHz is attributed to impedance mismatch between the electrode pads and the RF probe. The device with graphene coverage length of 19\,$\mu$m, see Fig.~\ref{fig:bandwidth} (D), exhibits a 1.5-dB optical bandwidth of 110\,GHz. 
With the increasing of the graphene coverage length, the responsivity is significantly improved as demonstrated in Fig.~\ref{fig:responsivity}(A), while the current frequency response $R_{\rm op}$ shown by the red triangular marks at the high frequency slightly drops, indicating a tradeoff between responsivity and bandwidth.

The detector with graphene coverage length of 19\,$\mu$m was further used to receive 10, 20, and 40\,Gbit/s return-to-zero (RZ) optical signals with pseudo-random binary sequence (PRBS) length of $2^{31}-1$, which is amplified by an erbium-doped fiber amplifier (EDFA) and injected to the chip. 
The output electrical signal from the graphene photodetector was electrically amplified (40\,GHz bandwidth electrical amplifier) and fed to the high-speed oscilloscope (70\,GHz bandwidth) to record the eye-diagram. Clear and open eye-diagrams were obtained for all 10, 20, and 40\,Gbit/s signals, as exhibited in Figs.~\ref{fig:bandwidth}(E)--(G), indicating no pattern effect of the detector and proving the feasibility of using such graphene photodetectors in realistic optical communication applications.

\subsection*{Discussion}

Table~\ref{SOTA} summarizes the state-of-the-art waveguide-coupled integrated graphene photodetectors, as well as a commercially available photodetector. The performance of more than 110\,GHz bandwidth with high responsivity of 360\,mA/W is significantly beyond previous graphene silicon waveguide photodetectors and is comparable to the state-of-the-art commercial high-speed photodetectors~\cite{XPDV412xR}.
With narrower plasmonic gap as well as thinner Au thickness, higher absorption by graphene is expected, thus promising even higher responsivity, as analyzed in the Supplementary Material. Furthermore, tuning the Fermi-level by top-gate~\cite{CL2014} would further improve the absorption of graphene and thus responsivity. 
Moreover, the use of CVD-growth graphene enables large scale integration. 
It should be noted that the narrow plasmonic slot of 120\,nm results in an ultra-fast transit of carriers to the electrodes. The transit-time-limited bandwidth of the photodetector is given by
$f_t=3.5/2\pi{t_{\rm tr}}$~\cite{FX2009}, where $t_{\rm tr}$ is the transit time through the photodetection region. At the zero bias voltage, the difference in Fermi level between the palladium- and titanium-doped graphenes is 0.05\,V~\cite{TM2010}, resulting in the built-in electric-field of $\sim$0.4\,V/$\mu$m. With a corresponding carrier velocity of 1.1$\times 10^5$\,cm/s~\cite{VD2010,IM2008,MB2017}, it takes only 1.1\,ps for the carriers to transit through the 120\,nm plasmonic slot gap.
Thus, a single transit-time limited bandwidth of 520\,GHz is expected, 
much larger than the measured bandwidth of 110\,GHz. It implies that there are possibilities to further optimize the performance of the graphene photodetector. 

\begin{table}[ht!]
\centering
\caption{Summary of the state-of-the-art waveguide-coupled integrated graphene PDs and commercial(*) semiconductor high-speed PD.}
\begin{tabular}{ c | c | c | c | c}
\hline \hline \noalign{\smallskip}
Reference & Responsivity & EQE & Bandwidth & Type of \\
&  &  &  & graphene \\ \hline   \noalign{\smallskip}
\cite{XG2013} &  100\,mA/W & 7.9$\%$ & 20\,GHz & Exfoliation \\ 
\cite{DS2014} &  7\,mA/W & 0.56$\%$ & 41\,GHz & CVD \\ 
\cite{NY2014} &  57\,mA/W & 4.56$\%$ & 3\,GHz & CVD \\ 
\cite{RJS2015} &  360\,mA/W & 29$\%$ & 42\,GHz & Exfoliation \\ 
\cite{SS2016} &  76\,mA/W & 6.4$\%$ & 65\,GHz & Exfoliation \\ 
\cite{IG2016} &  370\,mA/W & 29.5$\%$ & ---- & CVD \\ 
\cite{DS2017} &  1\,mA/W & 0.08$\%$ & 76\,GHz & CVD \\ 
(*)\cite{XPDV412xR} &  500\,mA/W & 40$\%$ & 100\,GHz & ----\\ 
This work &  360\,mA/W & 29$\%$ & >110\,GHz & CVD \\ 
\hline \hline \noalign{\smallskip}
\end{tabular}
\label{SOTA}
\end{table} 

In summary, we have demonstrated an on-chip ultra-high bandwidth photodetector based on a single-layer CVD graphene and plasmonic slot waveguide hybrid structure. The narrow plasmonic slot waveguide not only enhances light-graphene interactions, but also enables to separate the photo-generated carriers effectively, leading to 
high responsivity and large bandwidth. An optical bandwidth larger than 110\,GHz with large responsivity of 360\,mA/W is achieved. 
The devices demonstrated here are fully CMOS compatible and can easily be integrated with silicon platform, and the use of CVD-growth graphene paves a promising way to achieve multi-functional integrated graphene devices for optical interconnects.

\section*{Method} 
\vspace{-2mm}
\subsection*{Fabrication process} 
The device was fabricated on a commercial silicon-on-insulator sample with top silicon layer of 250\,nm and buried oxide layer of 3\,$\mu$m. The top silicon layer was first thinned down to 100\,nm by dry-etching (STS Advanced-Silicon-Etching machine, ASE) in order to obtain a good coupling efficiency with the plasmonic slot waveguide. The grating couplers and silicon waveguides were patterned by electron-beam lithography (EBL, E-Beam Writer JBX-9500FSZ), and full-etched by ASE dry-etching process. After that, the graphene layer was wet-transfered and patterned by standard ultraviolet (UV) lithography (Aligner: MA6-2) and oxygen (O$_2$) plasmonic etching. Then, the Au/Pd (90\,nm/3\,nm)-graphene contact was patterned by a second EBL, and obtained by metal deposition and lift-off process. Finally, the Au/Ti (90\,nm/3\,nm)-graphene contact was patterned by a third EBL, and obtained by metal deposition and lift-off process.

\vspace{-2mm}
\subsection*{Graphene wet-transferring process} 
Single-layer graphene grown by CVD on Cu foil (GRAPHENE SUPERMARKET) was transferred onto the devices by a wet chemistry method. The SLG sheet was transferred by the following steps. Firstly, a layer of photo-resist (AZ5200 series) was spin-coated (3500\,rpm for 1 minute) onto the SLG by a spin-coater and the photo-resist/SLG stack was then released by etching the underlying Cu foil in a Fe(NO3)3 solution (17 wt\%) at room temperature. The stack floating in the solution was then washed by deionized (DI) water for a couple of times and transferred to the target device followed by drying at room temperature for at least 24 hours. Afterwards, the photo resist on the SLG was dissolved in acetone at room temperature.  Finally, the graphene device was cleaned by ethanol and DI water and then dried before further processing.

\vspace{-2mm}
\section*{Acknowledgments} 
The author would like to thank Vitaliy Zhurbenko for the help of calibrating the RF cables and Bias Tee, and thank Prof. Fengnian Xia and Dr. Peter David Girouard for constructive discussion. The work is supported by the Center for Silicon Photonics for Optical Communication (SPOC, DNRF123), the Center for Nanostructured Graphene (CNG, DNRF103) both sponsored by the Danish National Research Foundation, and the mid-chip project sponsored by VILLUM FONDEN (No. 13367). N. A. M. is a VILLUM Investigator supported by VILLUM FONDEN (No. 16498) and X. Z. is supported by VILLUM Experiment (No. 17400).

\vspace{-1mm}
\section*{Supplementary Materials}
\vspace{-1mm}

Supplementary materials are available under request.





\end{document}